\documentclass[aps,prx,superscriptaddress,twocolumn]{revtex4-1}
\usepackage{amsmath}
\usepackage{amssymb}
\usepackage{graphics}
\usepackage{graphicx}
\usepackage{color}
\usepackage{hyperref}
\usepackage{bm}
\usepackage{txfonts}

\newcommand{\be}{\begin{equation}}
\newcommand{\ee}{\end{equation}}
\newcommand{\bk}{{\bm{k}}}

\begin{document}
	\title{Magnetoconductivity of type-II Weyl semimetals}
	\author{Yi-Wen Wei}
	\affiliation{International Center for Quantum Materials, School of Physics, Peking University, Beijing 100871, China}
	\affiliation{Collaborative Innovation Center of Quantum Matter, Beijing 100871, China}
	\author{Chao-Kai Li}
	\affiliation{International Center for Quantum Materials, School of Physics, Peking University, Beijing 100871, China}
	\affiliation{Collaborative Innovation Center of Quantum Matter, Beijing 100871, China}
	\author{Jingshan Qi}
	\affiliation{School of Physics and Electronic Engineering, Jiangsu Normal University, Xuzhou 221116, China}
	\author{Ji Feng}
	\email{jfeng11@pku.edu.cn}
	\affiliation{International Center for Quantum Materials, School of Physics, Peking University, Beijing 100871, China}
	\affiliation{Collaborative Innovation Center of Quantum Matter, Beijing 100871, China}
	
	\begin{abstract}
		Type-II Weyl semimetals are characterized by the tilted linear dispersion in the low-energy excitations, mimicking Weyl fermions but with manifest violation of the Lorentz invariance, which has intriguing quantum transport properties. The magnetoconductivity of type-II Weyl semimetals is investigated numerically based on lattice models in parallel electric and magnetic field. We show that in the high-field regime, the sign of the magnetoconductivity of an inversion-symmetry-breaking type-II Weyl semimetals depends on the direction of the magnetic field, whereas in the weak field regime, positive magnetoconductivity is always obtained regardless of magnetic field direction. We find that the weak localization is sensitive to the spatial extent of impurity potential. In  time-reversal symmetry breaking type-II Weyl semimetals, the system displays either positive or negative magnetoconductivity along the direction of band tilting, owing to the associated effect of group velocity, Berry curvature and the magnetic field.
	\end{abstract}
	
	\date{\today}
	\maketitle
	
	\section{Introduction}
	A Weyl semimetal  \cite{wan2011, Burkov2011, Burkov2011b, Xu2011, Wang2012, Wang2013, Weng2015, Xu2015, Lv2015}  hosts linear energy dispersions through the Weyl points in the electronic band structure and displays interesting quantum properties, such as Fermi-arc surface states and the chiral anomaly \cite{Adler1969, Bell1969, Nielsen1983}. As a manifestation of the chiral anomaly associated with a Weyl Fermion, positive magnetoconductivity \cite{Son2013} has been observed in Weyl semimetals \cite{Zhang2016, Huang2015, Arnold2015}. Recently, type-II Weyl semimetals were theoretically proposed and soon realized in experiments \cite{Weyl2015,Deng2016,Wang2016MoTe2,Wu2016, Huang2016,Tamai2016}. Unlike the usual, or type-I, Weyl semimetals, the linear band dispersion near a Weyl node in type-II Weyl semimetals is significantly tilted, so that the Fermi surface encloses both electron and hole pockets. It is therefore important to examine the consequence of tilted Weyl-type dispersion from a microscopic viewpoint in order to understand the extraordinary properties of type-II Weyl semimetals \cite{Yao2016,Tchoumakov2016,Udagawa2016}. An aim of the present work is to investigate theoretically the transport properties of type-II Weyl semimetals based on tight-binding models, with which the full extent of the band tilting in the k-space can be accessed. Both inversion-breaking type-II Weyl semimetals and time-reversal-breaking type-II Weyl semimetals are addressed in our calculations.
	
	Indeed, as a result of the tilted dispersion, the chiral anomaly and likewise the positive magnetoconductivity are believed to be absent in some directions \cite{Weyl2015,Yao2016,Tchoumakov2016,Udagawa2016}. It is also intriguing to notice that the observed magnetoconductivity in type-II Weyl semimetals \cite{Wang2016,Lv2017a} shows both direction- as well as sample-dependences. Evidently, this suggests that the nature of impurity and localization also play crucial roles in the transport properties of type-II Weyl semimetals, which is also the case for type-I Weyl semimetals \cite{Goswami2015, Lu2015b}. For type-II Weyl semimetals, the previous theoretical analyses are mainly based on the semi-classical Boltzmann approach \cite{Weyl2015, Sharma2016}. Detailed analyses on the quantum transport and the effects of the impurity potential are still lacking. Therefore, a second aim of this work is then to systematically investigate the conductivity of type-II Weyl semimetals under magnetic field, combining a quantum mechanical linear-response theory based on the Kubo formula and a semi-classical approach based on the Boltzmann equation. In this approach, the effects of the range of impurity potentials and the quantum mechanical interference in the transport can be quantitatively analyzed with the tight-binding models.
	
	We show that the Drude magnetoconductivity of the inversion-breaking type-II Weyl semimetals is positive in the weak magnetic field limit, while in the high field regime its sign is dependent on the magnetic field direction. These results are consistent with the recent experiments on $\text {WTe}_2$ \cite{Wang2016,Lv2017a}, and may also be justified theoretically. The quantum correction to the magnetoconductivity is found to be negative due to weak localization effect \cite{Abrahams1979, BERGMANN1984}, and its magnitude decreases with an increasing magnetic field, indicating the suppression of weak localization. In particular, we demonstrate that the weak localization in type-II Weyl semimetals decreases with increasing spatial extent of impurity potential. Finally, a tight-binding model without time-reversal symmetry is analyzed. We find that the magnetoconductivity can be either positive or negative along the band tilting direction, which depends on the combined effect of group velocity, Berry curvature and the direction of the magnetic field.
	
	The paper is organized as follows. Section \ref{S-Drude} demonstrates the magnetoconductivity of the inversion-breaking type-II Weyl semimetals, where the chiral anomaly and quantum oscillations are discussed. This is followed by an analysis of the quantum correction to magnetoconductivity in Section \ref{S-correction}. The magnetoconductivity of the time-reversal breaking type-II Weyl semimetals is studied in Section \ref{S-time}. The key results are summarized with an outlook is presented in Section \ref{S-sum}.
	
	\section{Drude conductivity of inversion-breaking type-II Weyl semimetals}\label{S-Drude}
\begin{figure}
		\centering
		\includegraphics[width=72 mm]{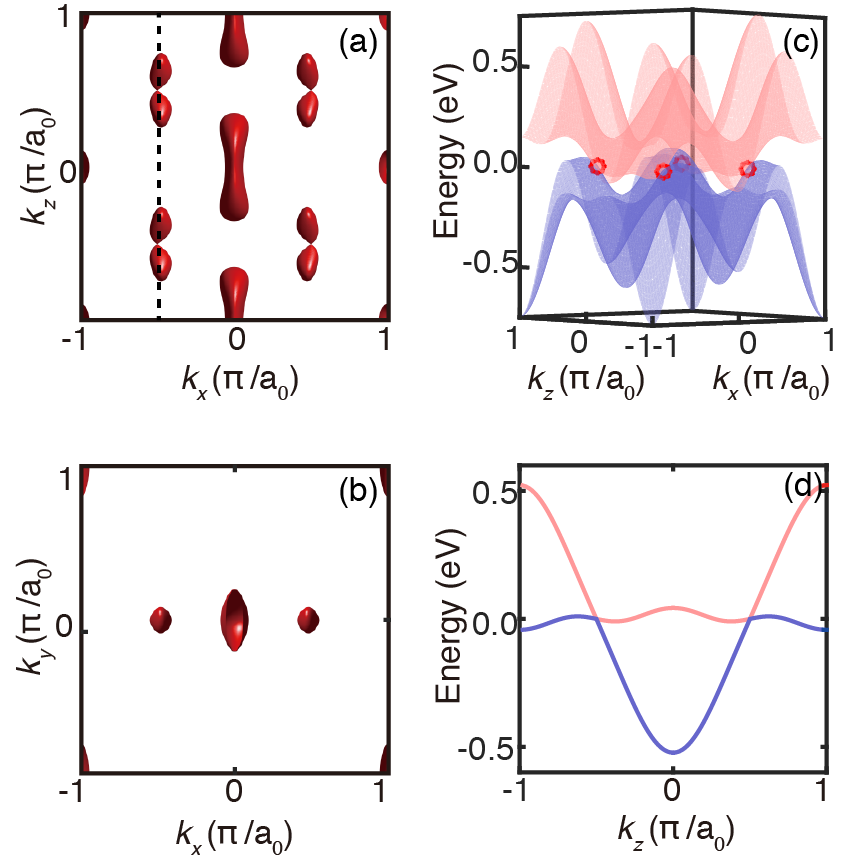}\\
		\caption{Fermi surfaces and band structures of the inversion-breaking model with parameters $t_x=t/2$, $m=2t$, $\gamma=2.4t$, $t=0.1$ eV, $k_0=\pi/2$ in Eq.~(\ref{eq:001}) and lattice constant $a_0=6$ {\AA}. Fermi surfaces with $E_F=0$ eV projected to (a) $k_x-k_z$ plane and (b) $k_x-k_y$ plane. (c) Band structures with $k_y=0$. The Weyl points are enclosed by red circles. (d) Energy spectrum through the Weyl nodes plotted along $k_z$ direction (as the dashed line in (a)).}
		\label{fig:fermi}
	\end{figure}

	\emph{Model and methods.---}
	When either inversion or time-reversal symmetry is broken, a Dirac node can be split into a pair of Weyl nodes that carry opposite chiralities \cite{Nielsen1981}. In this Section, we analyze the Drude conductivity of type-II Weyl fermions based on a minimal tight-binding model, which breaks inversion symmetry but keeps time-reversal symmetry \cite{McCormick2016}.  The two-band Hamiltonian without a magnetic field is given by
	\be
	H= c^\dagger_{\bm k\alpha}h^i\sigma^i_{\alpha\beta}c_{\bm k\beta},
	\label{eq:001}
	\ee
	in which $i =0,1,2,3$ and repeated indices are summed over. Here, $\sigma^0$ is $2\times2$ order identity matrix and $\sigma^1$, $\sigma^2$, $\sigma^3$ are Pauli matrices. $c_{\bm k\alpha(\beta)}^{\dagger}$ creates an electron with momentum $\bm k$ in orbital $\alpha(\beta)$. And
	\be
	\begin{split}
		&h^0=\gamma(\text{cos}2k_x-\text{cos}k_0)(\text{cos}k_z-\text{cos}k_0),\\
		&h^1=-m(1-\text{cos}^2 k_z-\text{cos}k_y)-2t_x(\text{cos}k_x-\text{cos}k_0),\\
		&h^2=-2t\text{sin}k_y, \quad h^3=-2t\text{cos}k_z.
	\end{split}
	\ee
	The Fermi surfaces of $E_F=0$ eV and the band structures are shown in Fig.~\ref{fig:fermi}. At the Fermi level, there are four electron-hole pockets corresponding to the four Weyl nodes at $(\pm k_0,0,\pm k_0)$, with $k_0 = \pi/2$ in our calculations. Meanwhile, it is seen that there exist trivial Fermi pockets  near $(0,0,0),(0,0,\pi),(\pi,\pi,0)$, and $(\pi,\pi,\pi)$, which are far away from the Weyl nodes. The dispersion of the Weyl node is tilted along the $z$-direction, as depicted in Fig.~\ref{fig:fermi} (d).
	
	The modification of hopping amplitude by the presence of impurities on sites $R_i$ can be modeled by
	\be
	H'=\sum_{\bm R,\bm R_i,\alpha} V(\bm R-\bm R_i) c_{\bm R \alpha}^{\dagger} c_{\bm R \alpha},
	\label{eq:003}
	\ee
	where $V(\bm R-\bm R_i)$ stands for the impurity potential at site $\bm R$ produced by the impurity site $\bm R_i$. In this Section, short-ranged impurity potential $V(\bm R-\bm R_i)= u\delta_{\bm R,\bm R_i}$ is employed. In Section \ref{S-correction}, $V(\bm R - \bm R')$ with finite spatial extent will be examined. The impurity sites are assumed to be randomly and uniformly distributed on the lattice, with a concentration $n$. The self-consistent Born approximation is used to compute electron lifetimes. Subsequently, the Kubo formula will be used to calculate the Drude conductivity, in which the magnetic field enters into the Hamiltonian in Eq. (\ref{eq:001}) via the standard Peierls substitution. In the weak magnetic field limit, the semi-classical Boltzmann equation is used in place of the Kubo formula, as the magnetic supercells required by the Peierls substitution become excessively large. Further details of the model and the computational method can be found in Appendix~\ref{appendix}.
	
	\emph{Drude Conductivity along $x$/$z$ direction.---}
	For the model under scrutiny the band dispersion across the Weyl nodes along the $z$-direction (tilting direction) is intrinsically different from that along the $x$-direction. Correspondingly, the responses of conductivity to magnetic field along $x$- and $z$-direction are expected to be distinct. The magnetic field to be considered takes the form $\bm B=(B_x,0,B_z)$ with $B_x=0$ or $B_z=0$ corresponding to magnetic field along $z$- and $x$-direction respectively. The corresponding vector potential $\bm A=(-yB_z,0,yB_x)$ breaks translational symmetry along $y$-direction, which is taken into account by the Peierls substitution implemented with a $1 \times L_y \times 1$ magnetic supercell. $L_y$ is related to the magnetic field by $L_y=2\pi/B$.
	\begin{figure}
		\centering
		\includegraphics[width=68mm]{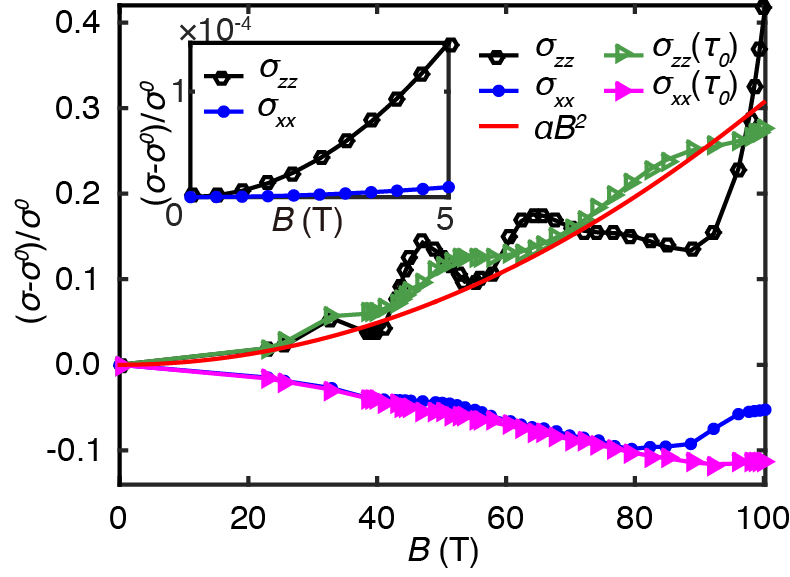}\\
		\caption{ Drude conductivity under high magnetic field with $\bm B\varparallel\bm E$.  $\sigma_{zz}$ and $\sigma_{xx}$ stand for the  conductivity along $z$- and $x$-direction, respectively. The red curve is the fitted result by $\alpha B^2$ with $\alpha=4.046\times10^{-5}$. $\sigma_{zz}(\tau_0)$ and $\sigma_{xx}(\tau_0)$ are obtained with fixed relaxation time $\tau_0$, $\sigma^0$ is the conductivity under zero magnetic field. Inset shows the conductivity under relatively weak magnetic field using the semi-classical Boltzmann equation.}
		\label{fig:drude}
	\end{figure}	
	The calculated Drude conductivity always decreases with increasing $B$ when the magnetic field $\bm B$ is perpendicular to the electric field $\bm E$, owing to the deflection of electron trajectories by the Lorentz force. We shall thus focus on the case $\bm B\varparallel\bm E$ in this work. The computed conductivity in this geometry of external fields is plotted as a function of $B$ in Fig.~\ref{fig:drude} with $u^2n=0.002$ eV$^2$ and $E_F=0$ eV. The conductivity along $z$-direction, $\sigma_{zz}$, increases  with increasing $B$ in an oscillatory fashion. If we ignore the oscillations in $\sigma_{zz}$, the calculated positive magnetoconductivity along $z$-direction displays a quadratic dependence on $ B$ with $\alpha  B^2$, as indicated by the red curve in Fig.~\ref{fig:drude}. This is expected for magnetoconductivity induced by chiral anomaly. It is noted that this positive magnetoconductivity from chiral anomaly is unaffected by the trivial Fermi pockets. In contrast to the case of $\sigma_{zz}$, the conductivity along $x$-direction $\sigma_{xx}$ decreases with increasing $B$ when 25 T $<B<$ 75 T. In the relatively weak field limit the magnetoconductivity, obtained using the semi-classical Boltzmann equation, appears  to be positive in both $x$- and $z$-direction, as displayed in the inset of Fig.~\ref{fig:drude}. We have also examined the case when the Fermi level deviates slightly from half-filling, and found that the magnetoconductivity is qualitatively similar to the half-filling case ($E_F = 0$ eV), as described above.	
	
	It has been demonstrated with a semi-classical approach that the chiral anomaly of type-I Weyl fermions can lead to a positive magnetoconductivity quadratic in $B$ in the presence of parallel electric and magnetic fields\cite{Son2013}. In contrast to type-I Weyl fermions, the general trend of the magnetoconductivity of type-II Weyl fermions is separated into two regimes, as indicated by our calculations. In the quantum regime (high field), the chiral anomaly induced positive magnetoconductivity emerges only if the discrete Landau levels are formed under the given magnetic field. However, in the low-field regime the positive magnetoconductivity is shown to  exist  when $\bm B$ is along any arbitrary direction based on the Boltzmann approach, as shown in the inset of Fig. \ref{fig:drude}. These two different behaviors  provide a microscopic reconciliation of two transport experiments in $\text{WTe}_2$, where our calculated $\sigma_{zz}$ and $\sigma_{xx}$ in high field ($B>25$ T) and weak field limit conform to the two regimes, respectively\cite{Wang2016, Lv2017a,Sharma2016}.
	
	\begin{figure}
		\centering
		\includegraphics[width=80mm]{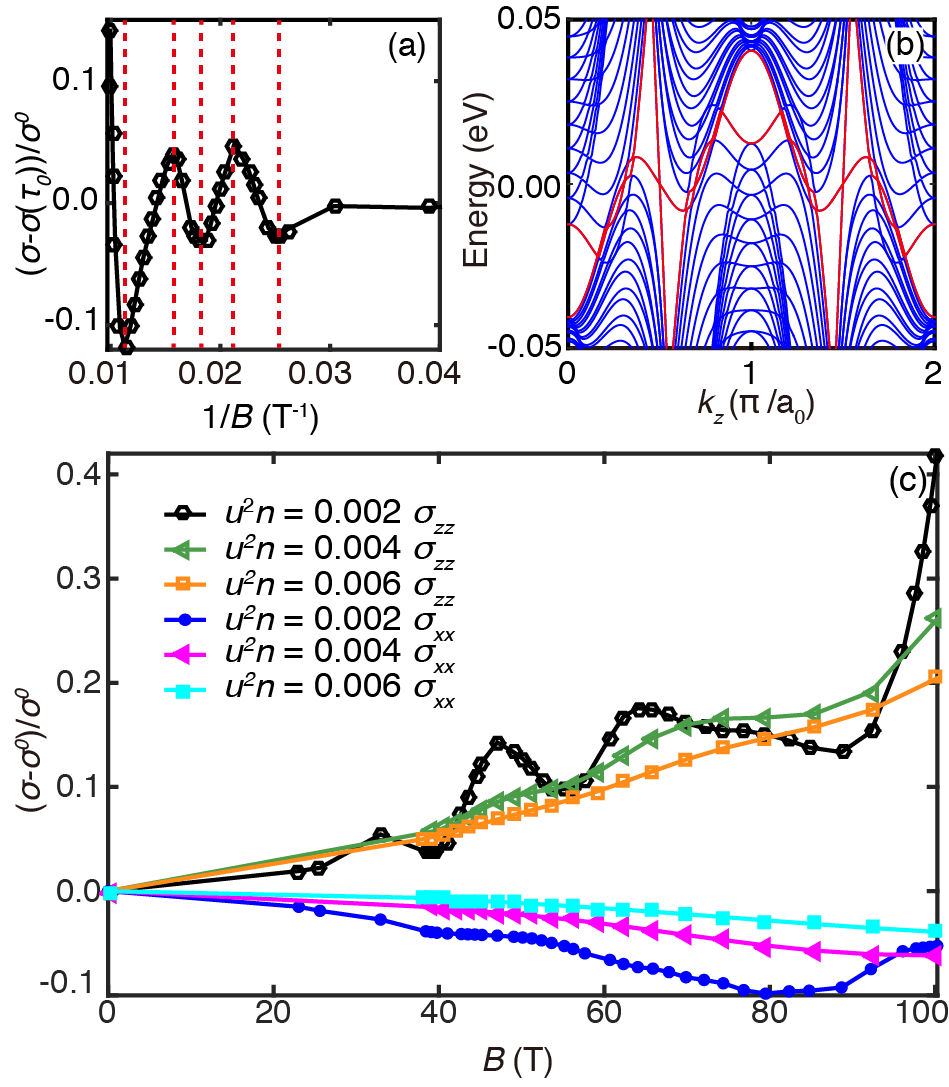}\\
		\caption{Quantum oscillations of the  magnetoconductivity. (a) Quantum oscillations of the relative conductivity along $z$-direction as function of $1/B$. (b) The energy spectra along $k_z$ direction with the magnetic field $B=50$ T and the Fermi energy $E_F=0$ eV. (c) Quantum oscillations with different impurity strengths measured by $u^2n$. The legends $\sigma_{zz}$ and $\sigma_{xx}$ mark the  conductivity in $z$- and $x$-direction, respectively.}
		\label{fig:sdh}
	\end{figure}
	
	\emph{Shubnikov-de Haas oscillations.---}
	The oscillations in $\sigma_{zz}$ shown in Fig.~\ref{fig:drude} is attributable to the Shubnikov-de Haas (SdH) oscillations arising from the oscillation of relaxation time, which in turn reflects the oscillation of spectra as the magnetic field increases. A confirmation can be obtained by computing the magnetoconductivity  with the relaxation time fixed to the value $\tau_0$ under zero magnetic field instead of that derived from the self-consistent Born approximation with actual magnetic field. The results indicate that $\sigma_{zz}(\tau_0)$ increases without any obvious oscillations as $B$ increases. In order to focus on the oscillations, we remove the effect of chiral anomaly by considering a relative conductivity $(\sigma_{zz}-\sigma_{zz}(\tau_0))/\sigma^0_{zz}$, where $\sigma^0_{zz}$ is the conductivity under $B=0$. The relative conductivity is presented as a function of $1/ B$ in Fig.~\ref{fig:sdh} (a). According to the Lifshitz-Kosevich formula \cite{lifshitz1956theory}, the conductivity with oscillations is periodic on $1/B$ with a relation of $\text{cos}(2\pi(F/B+\phi))$, where $F$ is the frequency and $\phi$ is the phase shift. However, the calculated oscillations do not exhibit single periodicity to be fitted with the Lifshitz-Kosevich formula, possibly due to the complicated Fermi surfaces as shown in Fig.~\ref{fig:sdh} (b). Indeed, the experiments have observed the SdH oscillations with multiple frequencies in Weyl semimetals hosting multiple Fermi pockets \cite{Hu2016, Wang, Huang2015}.
	
	Furthermore, it is seen that the amplitude of the SdH oscillations in $\sigma_{zz}$  becomes reduced as the impurity strength measured by $u^2n$ increases. For sufficiently large values of $u^2n$, the SdH oscillations become completely suppressed, as demonstrated in Fig.~\ref{fig:sdh} (c). This reduction and eventual disappearance of the SdH oscillations  arises from the decrease of relaxation time at large $u^2n$, which  in turn leads to increasing broadening of the Landau levels. This implies that the spectral oscillation that causes SdH oscillation gets less sharp as the impurity scattering is increased, resulting in weaker SdH oscillations. We expect that the increase of interaction range, resulting in smaller relaxation time, could also suppress the SdH oscillations. These results and analyses provide an explanation to the experimental observation that the SdH oscillations displayed sample-dependent features at the same temperature and the same magnetic field\cite{Zhang2016}. Similarly, in experiment the rising temperature can bring a thermal effect to broaden the Landau levels and thus suppress the SdH oscillations, which has already been observed in experiment\cite{Zhang2016}. Lastly, we note that the calculated conductivity along $x$-direction is always reduced by increasing magnetic field.		
	
	\section{Quantum correction to conductivity} \label{S-correction}
	\begin{figure}
		\centering
		\includegraphics[width=70mm]{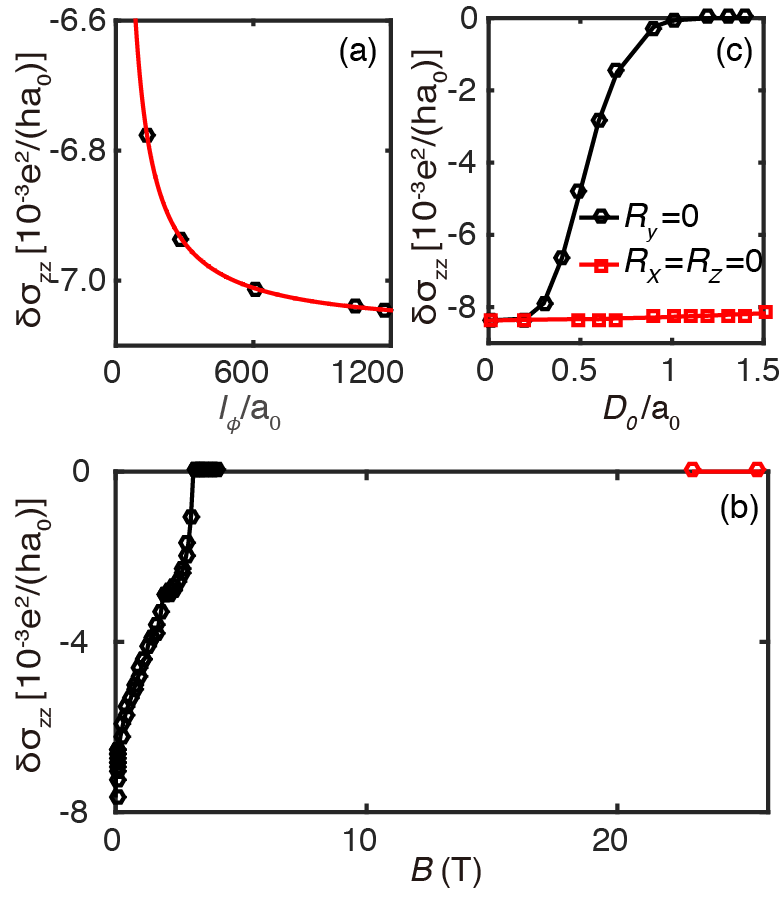}\\
		\caption{ Computed quantum correction to the conductivity. (a) $\delta\sigma_{zz}$ as a function of coherent length $l_\phi$. $a_0$ is the lattice constant. The localized potential is applied with the impurity potential and concentration satisfying $u^2n=0.002$ $\text{eV}^2$. (b) $\delta\sigma_{zz}$ as a function of $ B$, where the values of $\delta\sigma_{zz}$ are computed by Kubo formula employing magnetic supercell in high-field regime (red polygons) and by using the semi-classical approximation in low-field regime (black polygons).  (c) Interaction range ($D_0$) dependence of the quantum correction to the conductivity with $l_\phi=1177a_0$, where the mean-free path $l=6.99a_0$ is determined by relaxation time and diffusion coefficient.}
		\label{fig:quantum}
	\end{figure}
	The Drude conductivity presented above does not include the so-called  weak (anti-)localization \cite{Abrahams1979, BERGMANN1984}, while this type of quantum correction to the conductivity could play an important role in the transport especially in  time-reversal invariant systems at low temperatures. When scattering mechanism is being considered, the inherent anisotropy in type-II Weyl fermions leads to rather different symmetry classification from the isotropic Weyl fermions\cite{Lu2015} . Consequently, the quantum correction to the Drude conductivity warrants careful scrutiny. This is achieved by solving the 4-point Dyson equation corresponding to the maximally-crossed diagrams \cite{Neal1966}, which is the leading order contribution to the quantum correction from the diagrammatic averaging of the Kubo formula over the ensemble of disorder configurations.
	
	In Fig.~\ref{fig:quantum} (a), we present the computed quantum correction to conductivity ($\delta\sigma_{zz}$) of the tight-binding model in Eq. (\ref{eq:001}) along with a localized impurity potential, under zero magnetic field as a function of $l_\phi$. $l_\phi$ is the coherence length in $z$-direction charactering the inelastic scattering process, as shown in Appendix~\ref{appendix}. It is clear that $\delta\sigma_{zz}$ is negative, indicating weak localization. The weak localization correction to conductivity in  anisotropic system theoretically takes the form\cite{wolfle1984electron}
	\be
	\delta\sigma_{zz}=\frac{e^2}{h\pi^2}\alpha(1/l_{\phi}-1/l),\label{eq:lphi}
	\ee
	with $l$ denoting the mean-free path and $\alpha=\frac{\sigma_{zz}}{\sqrt{\sigma_{xx}\sigma_{yy}}}$. The effects of anisotropy are absorbed into the anisotropic coefficient $\alpha$. Fitting the calculated $\delta\sigma_{zz}$ versus $l_\phi$ shown in Fig.~\ref{fig:quantum} (a) to the formula in Eq. (\ref{eq:lphi}) gives the fitted values $\alpha=0.41$ and $l=5.85a_0$, which are quite different from $\alpha$ given by $\frac{\sigma_{zz}}{\sqrt{\sigma_{xx}\sigma_{yy}}}=0.43$ and the mean-free path $l=8.89a_0$ determined by the relaxation time and the diffusion coefficient (see Appendix).

\begin{figure}
		\centering
		\includegraphics[width=75mm]{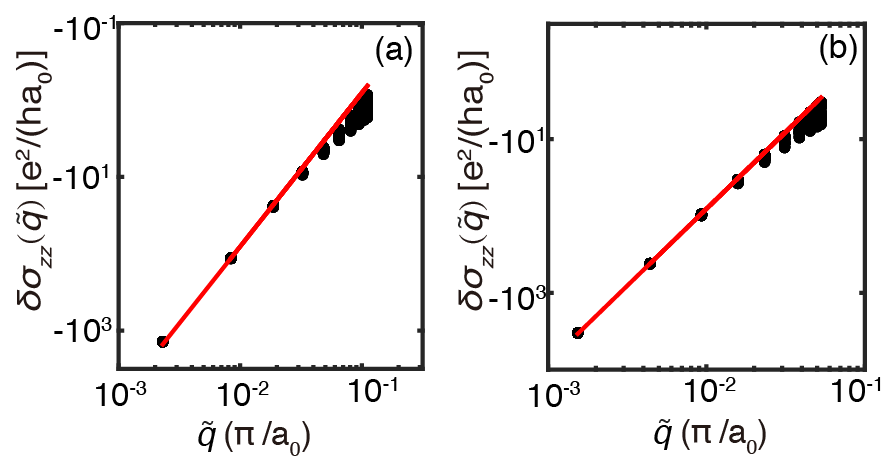}\\
		\caption{Quantum correction to conductivity $\delta\sigma_{zz}(\tilde {\bm q})$  versus $\tilde{q}$. $\delta\sigma_{zz}(\tilde {\bm q})$ as a function of $\tilde{q}$ for (a) type-II Weyl fermions as $E_F=0$ eV with parameters of Hamiltonian in Eq.~(\ref{eq:001})  the same as in Fig.~\ref{fig:fermi} and for (b) type-I Weyl fermions as $E_F=$ 0.15 eV with parameters $\gamma$ and $t_x$  in Eq. (\ref{eq:001}) changed to $0$ and $t$  respectively. The plots are shown in logarithmic coordinate and the red lines show the results given by $\delta\sigma_{zz}(\tilde{\bm q})=-e^2/(4\pi^3h\tilde{q}^2)$. The short-ranged impurity potential is employed with $u^2n=0.002$eV$^2$.
		}\label{fig:correction-f}	
	\end{figure}

	To understand the mismatch between the fitted and the theoretically derived values of $\alpha$ and $l$, it is useful to analyze $\bm{q}$-resolved conductivity change, $\delta\sigma_{zz}(\bm q)$, which when integrated yield the total quantum correction to conductivity. Under the assumption that the diffusion coefficient is anisotropic but $\bm q$-independent,
	$\sigma_{zz}(\bm q)$ has $\frac{-e^2}{4\pi^3h} (D_{xx}/D_{zz}q_x^2+D_{yy}/D_{zz}q_y^2+q_z^2)^{-1}$ near the diffusion pole($q =0$), where $D_{xx},D_{yy},D_{zz}$ are diffusion coefficients. Rescaling the momenta as $ \tilde{\bm q} \equiv(\sqrt{\frac{\sigma_{xx}}{\sigma_{zz}}}q_x, \sqrt{\frac{\sigma_{yy}}{\sigma_{zz}}}q_y, q_z)$, we have $\delta\sigma_{zz}(\tilde{\bm q})=-e^2/(4\pi^3h\tilde{q}^2) $, using the fact that the ratios of diffusion coefficients are equal to the ratios of conductivity via the Einstein relation. It is easy to find that  the integral of $\sigma_{zz}(\tilde{\bm q})$ exactly gives the relation shown in Eq. (\ref{eq:lphi}). The calculated $\delta\sigma_{zz}(\tilde{\bm q})$ versus $\tilde{q}$ is presented in Fig.~\ref{fig:correction-f} (a) in logarithmic coordinates. We find that when $\tilde{q}$ is small, the calculated  $\delta\sigma_{zz}(\tilde{\bm q})$ is almost the same for different $\bm q$ with the same length and can be well fitted by the formula $\delta\sigma_{zz}(\tilde{\bm q})=-e^2/(4\pi^3h\tilde{q}^2)$ describing the correction in the 3-dimensional isotropic systems. However,
	for larger $\tilde{q}$ the calculated $\delta\sigma_{zz}(\tilde{\bm q})$  deviates significantly from ${\tilde q}^{-2}$, and quite remarkably, are generally smaller than the corresponding values obtained by $\delta\sigma_{zz}(\tilde{\bm q})=-e^2/(4\pi^3h\tilde{q}^2)$. This makes the fitted $l$ smaller than the theoretical value. By changing the parameters $\gamma$  to 0 and $t_x$ to $t$ in Eq.~(\ref{eq:001}), a model of type-I Weyl fermions is obtained. The  anisotropy of $\delta\sigma_{zz}(\tilde{\bm q})$ is also found in type-I Weyl fermions as shown in Fig.~\ref{fig:correction-f} (b) with the localized impurity potential $u^2n=0.002$ eV$^2$ and $E_F=0.15$ eV.
	Thus our results and analysis indicate that the anisotropic efficient $\alpha$ can only capture the anisotropy from the scattering processes associated with the small total momentum. Therefore, for both type-I and -II Weyl fermions with anisotropy, the quantum correction $\delta\sigma_{zz} (\bm q)$ originating from the scattering processes with large total momentum, can have significant dependence on high-orders of $\bm q$; that is, the diffusion coefficients are no longer constants but now become $\bm q$-dependent.

	It is known that weak localization, reflecting quantum interference between time-reversed paths, can be suppressed by the application of an external magnetic field. To account for the external magnetic field, $\delta\sigma_{zz}$ is computed by the Kubo formula in magnetic supercells in the high-field regime, and with a semi-classical approximation (details in Appendix~\ref{appendix}) in low-field regime\cite{BERGMANN1984}, respectively. The computed values of $\delta\sigma_{zz}$ are shown as a function of magnetic field $B$ in Fig.~\ref{fig:quantum} (b). Indeed, as the addition of a magnetic field suppresses the weak localization, the conductivity increases quickly with increasing $B$. In fact, the quantum correction vanishes at around $B=3$ T from the semi-classical calculations, which is consistent with the result of Kubo formula by computing the maximally-crossed diagrams in supercells at the high-field limit.
	
	The spatial extent of impurity potential has an important influence on the quantum correction to conductivity \cite{Suzuura02}, which we now examine for the type-II Weyl fermion, by using the scattering Hamiltonian in Eq.~(\ref{eq:003}) with finite ranges of impurity potential. We adopt a Gaussian-type impurity potential in order to model the influence of spatial range on this quantum correction. The impurity potential is independent of orbital, and is expressed as
	\be
	\label{VR}
	V(\bm R-\bm R_i)=u\prod_{j}\frac{a_0}{\sqrt{2\pi}  R_0^j} e^{-(R^j- R_i^j)^2/(2{R_0^{j}}^2)},
	\ee
	where $R^j$ represents the component of $\bm R$ along $j$ direction ( $j=x, y, z$), and $\bm R_0$ characterizes the spatial ranges in the three directions.
	
	To focus on the quantum correction to conductivity contributed by the Weyl fermions, in evaluating the quantum correction from the maximally-crossed diagrams we only include the $\bm k$-points close to each Weyl point (distance between the k points to nearby Weyl point smaller than $\pi/4a_0$) so as to exclude the contributions from the trivial Fermi pockets. In this time-reversal invariant system, the quantum interference is dominated by intervalley scattering between states related by time-reversal symmetry. As shown in Fig. \ref{fig:fermi}, all four Weyl nodes are located on the $k_y=0$ plane and the extent of Fermi surfaces around each Weyl point are small in the $y$-direction. It is therefore expected that the quantum correction depends more sensitively on the spatial extent of impurity potential in the $x-z$ plane, than on the spatial extent along the $y$-direction. For this reason, it is sensible to devise the components of $\bm R_0$  such that $R_0^y \to 0$ and $R_0^x=R_0^z=D_0$ to reduce computational cost. The computed quantum correction to conductivity with the same $l_\phi$ and $l$ is seen to be suppressed by increasing the range $D_0$ of impurity potentials, and finally vanishes as soon as $D_0$ approaches the lattice constant $a_0$, as shown in Fig.~\ref{fig:quantum} (c). These results imply that the long-range potential could destroy the quantum interference between the scattering paths interrelated by time-reversal symmetry. This is expected for a time-reversal invariant system with a pair of valleys on the Fermi surface related by time-reversal symmetry. The quantum interference diminishes as the increase in spatial extent of the impurity potential reduces intervalley scattering. The case $R^y_0=D_0$ ($R^x_0$ and $R^z_0 \to 0$) is also computed for comparison, where the correction $\delta\sigma_{zz}$ does not change over $R^y_0$ as expected.
	
	\section{Drude conductivity of time-reversal-breaking type-II Weyl semimetals} \label{S-time}
	
	\begin{figure}
		\centering
		\includegraphics[width=72mm]{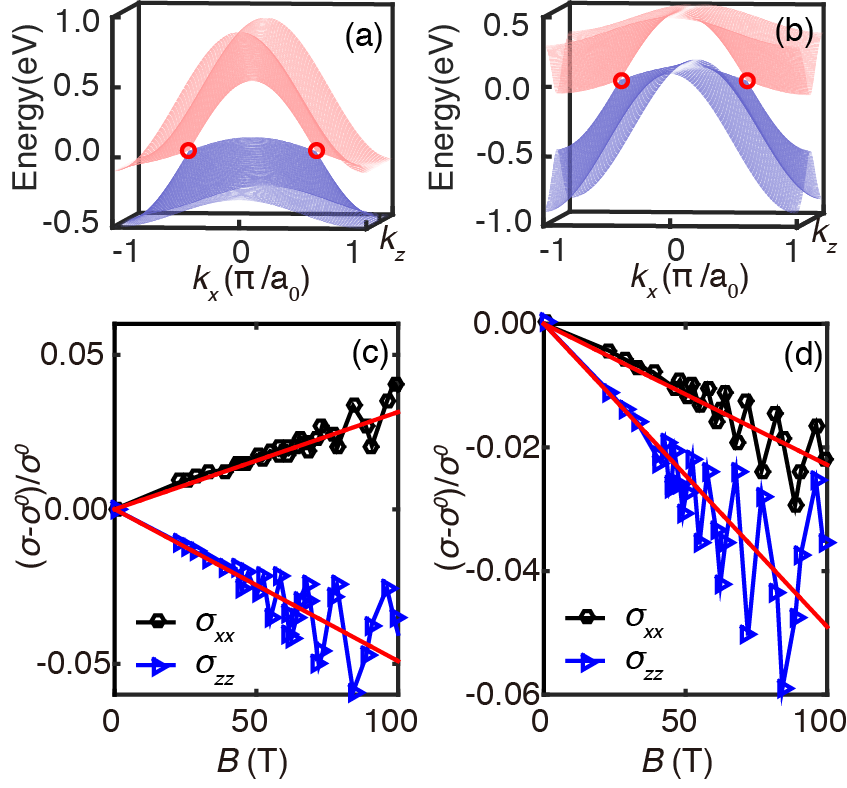}\\
		\caption{Band structures and magnetoconductivity of time-reversal breaking model when $\bm B$ is along the same direction of $\bm E$.
			Panels (a)--(b): Band structures of the Hamiltonian in Eq. (\ref{eq:002}) with $k_y=0$. The Weyl points are enclosed by red circles. We chose $m=2t$, $\gamma=3t$, $t=0.1$ eV, $k_0=\pi/2$, $a_0=6$ {\AA}, $E_F=0$ eV for (a) with $t_x=t$ and for (b) with $t_x=-t$. Magnetoconductivity shown in (c) and (d) correspond to the band structures in (a) and (b) respectively, where $\sigma_{xx}$ and $\sigma_{zz}$ represent the magnetoconductivity with the magnetic field applied along $x$- and $z$-direction respectively. The red lines are the fitted results with $\alpha  B$. The localized impurity potential satisfy $u^2n=0.002$ $\text{eV}^2$.}
		\label{fig:time}
	\end{figure}
	
	We now turn to a brief discussion of the magnetoconductivity of time-reversal-breaking type-II Weyl semimetal. If we consider the magnetoconductivity from a semi-classical viewpoint, as given in Eq. (\ref{eq:010}), it is seen that Berry curvature can make a crucial contribution. If a system breaks time-reversal symmetry, but preserves inversion symmetry, we have $\Omega_{\bk} = \Omega_{-\bk}$ and $\bm v_{\bk} = -\bm v_{-\bk}$. Upon the zone sum prescribed in Eq. (\ref{eq:010}), it clearly leads to a non-vanishing contribution to the magnetoconductivity first order in $B$. There is, however, no further constraint that determines the sign of the magnetoconductivity. Indeed, our calculations confirm that we can have a situation where the magnetoconductivity along the band tilting direction is either positive or negative in a time-reversal breaking type-II Weyl semimetals contrary to the conclusion drawn previously~\cite{Sharma2016}.
	
	The minimal model of time-reversal breaking type-II Weyl fermions takes the following form \cite{McCormick2016}
	\be
	H^T= c^\dagger_{\bm k\alpha}h_T^i\sigma^i_{\alpha\beta}c_{\bm k\beta},
	\label{eq:002}
	\ee
	in which
	\be
	\begin{split}
		&h_T^0=\gamma(\text{cos}k_x-\text{cos}k_0), h_T^2=-2t\text{sin}k_y, h_T^3=-2t\text{sin}k_z,\\
		&h_T^1=-m(2-\text{cos}k_y-\text{cos}k_z)-2t_x(\text{cos}k_x-\text{cos}k_0).\\
	\end{split}
	\ee
	This model hosts two Weyl nodes at $(\pm k_0, 0, 0)$ with $k_0=\pi/2$. The dispersion  of the Weyl nodes tilt in the $x$-direction. We shall examine two typical situations $t_x=t$ and $t_x=-t$, where the components of the Berry curvature around the Weyl points have opposite signs between these two cases, featuring opposite chiralities. The band structures are depicted in Fig.~\ref{fig:time} (a) and (b). Owing that the first-order terms of magnetoconductivity in $B$ are dependent on the sign of the magnetic field, both the parallel and anti-parallel electric and magnetic field are discussed.
	
	The Drude conductivity is calculated by Kubo formula with magnetic field applied  by Peierls substitution with the short-ranged impurity potential $u^2n=0.002$eV$^2$. The absence of time-reversal symmetry destroys the quantum interference between the scattering paths, making the quantum correction negligible, thus it is not included here. The Drude conductivity with $E_F = 0$ eV  is shown in Fig.~\ref{fig:time} (c) with $t_x=t$ and Fig.~\ref{fig:time} (d) with $t_x=-t$, where the magnetic field is applied along the same direction with the electric field. The oscillatory pattern of magnetoconductivity is once again a manifestation of the SdH oscillations. The value of $\sigma_{zz}$ decreases with increasing $B$ for both $t_x=t$ and $t_x=-t$. Besides, the value of $\sigma_{xx}$ increases linearly with $B$ for $t_x=t$, in accordance with previous work based on the Boltzmann approach\cite{Sharma2016}. However, when $t_x=-t$ the computed $\sigma_{xx}$ is negative for finite value of $B$.  In contrast to the case that $\bm B $ and $\bm E $ are in the same direction, the calculated $\sigma_{xx}$ become decreasing for $t_x=t$ and increasing for $t_x=-t$ with increasing $B$, when the magnetic field is applied in the opposite direction of the electric field. Actually, when $E_F = 0$ eV the magnetoconductivity for $t_x=t$ with $-\bm B$ exactly equals to the case for $t_x=-t$ with $\bm B$ according to the form of Hamiltonian.
	
\begin{figure}
		\centering
		\includegraphics[width=80mm]{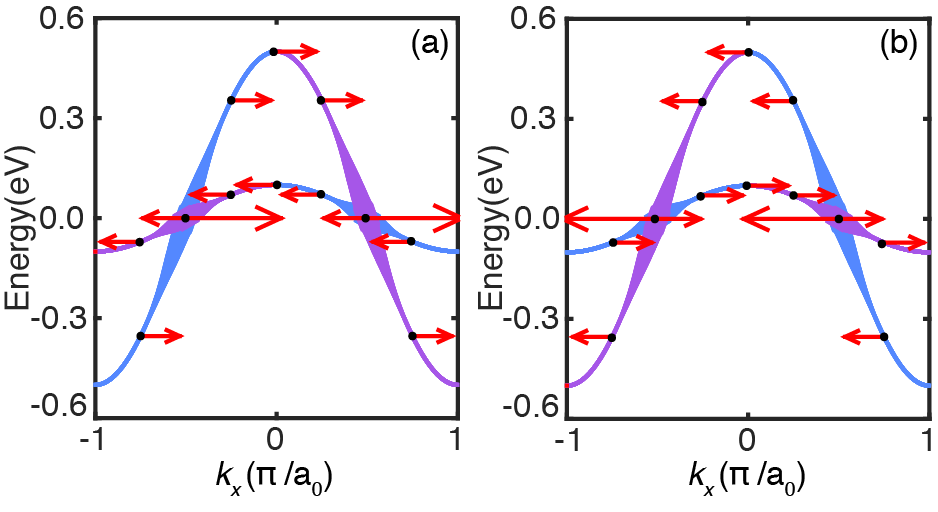}\\
		\caption{Depiction of $\Omega_{n\bm k}\cdot \bm v_{n\bm k}$ and  $v^x_{n\bm k}(\Omega_{n\bm k}\cdot \bm v_{n\bm k}) $ along ($k_x$, 0, 0), where (a) and (b) correspond to the $t_x=t$ and $t_x=-t$, respectively.
			The lines represent $\Omega_{n\bm k}\cdot \bm v_{n\bm k}$, where the line thickness represents the magnitude.  The purple (blue) color marks that the sign of $\Omega_{n\bm k}\cdot \bm v_{n\bm k}$ is positive (negative).  The corresponding $v^x_{n\bm k}(\Omega_{n\bm k}\cdot \bm v_{n\bm k})$ are represented by red arrows. The lengths of arrows correspond to the magnitude. }
		\label{fig:berry}
	\end{figure}

	To understand the results of $\sigma_{xx}$ from a perspective of semi-classical theory, we investigate the first-order terms of Eq.~(\ref{eq:010}),  $B^xv^x_{n\bm k}(\Omega_{n\bm k}\cdot \bm v_{n\bm k})$, in which the signs and magnitudes of $\Omega_{n\bm k}$ , $\bm v_{n\bm k}$ and $B^x$ are clearly all important. $\Omega_{n\bm k}\cdot \bm v_{n\bm k}$ near the Weyl nodes is seen to  dominate owing to the divergent $\Omega_{n\bm k}$, and has opposite signs for $\bm k$ and $-\bm k$ as shown in Fig.~\ref{fig:berry}.
	When multiplied by the factor $v_{n\bm k}^x$, the resulted quantity  $v^x_{n\bm k}(\Omega_{n\bm k}\cdot \bm v_{n\bm k})$ is dominated by the contributions from the Weyl nodes, which for $t_x=t$ is positive for the two Weyl nodes in Fig.~\ref{fig:berry} (a). This results in a positive overall magnetoconductivity along $x$-direction for $B^x>0$. In the case $t_x=-t$, the dominated first-order terms $v^x_{n\bm k}(\Omega_{n\bm k}\cdot \bm v_{n\bm k})$ are negative  as shown in Fig.~\ref{fig:berry} (b), and lead to positive  overall magnetoconductivity along $x$-direction for $B^x>0$. These semi-classical expectations agree very well with our numerical results. And it is  evaluated that the sign change of $B^x$ could also lead to the sign change of magnetoconductivity. These results quantitatively demonstrate that the magnetoconductivity of time-reversal breaking type-II Weyl fermions can be either positive ot negative along the band tilting direction, depending on the combined effect of group velocity, Berry curvature and the magnetic field.

	\section{Summary} \label{S-sum}
	
	To summarize, the magnetoconductivity of the type-II  Weyl semimetals is systematically analyzed based on tight-binding models, in which the effects of magnetic field, spatial extent of the impurity potential and quantum correction are taken into account. For the inversion-breaking type-II Weyl semimetal, the magnetoconductivity is always positive along the band tilting direction, whereas if the magnetic field is perpendicular to the band tilting direction such that chiral Landau levels are absent, the magnetoconductivity is negative under high magnetic field and positive under relatively weak magnetic field.
	The accompanying SdH oscillations are found not to exhibit simple periodicity  due to the complicated Fermi surfaces, and the oscillations can be suppressed by increasing the impurity potential or concentration. Taking  into account the quantum correction to conductivity, our results indicate that weak localization is present and it  should  be described by anisotropic and $\bm q$-dependent diffusion coefficients. It is found that the quantum correction increases with increasing magnetic field or increasing spatial extent of the impurity potential,  either of which suppresses weak localization. For the time-reversal-breaking type-II Weyl semimetals, the sign of magnetoconductivity can be either positive or negative along the band tilting direction, which is determined by an interplay of  group velocity, Berry curvature and the magnetic field near Weyl nodes.
	
	\section*{Acknowledgements}
	The work is supported by NSFC 11725415 and MOST 2016YFA0301004.
	
	\appendix
	\section{Computational details}\label{appendix}
	\emph{Drude conductivity.---}
	The impurities break translational symmetry, which result in scatterings between different k points. To restore the symmetry, we adopt the impurity configuration averaged Green's function with self-consistent Born approximation.
	\be
	\begin{split}
		&\bar{G}(\bk)=(E_F-H_{\bk}-\Sigma({\bk}))^{-1}\\
		&\Sigma(\bk)_{nm}=\sum_{\bm k^\prime}\sum_{n_1, n_2}\langle H'_{n\bk  n_1\bk'}H'_{n_2\bk' m\bk}\rangle \bar{G}_{n_1 n_2}(\bk'),
	\end{split}
	\ee
	where $\langle ...\rangle$ represents the average over impurity configurations and $m,n,n_1,n_2$ are the band indices. The imaginary part of $\Sigma(\bm k)_{nn}$ is directly related to the inverse of relaxation time by $\tau_{n\bm k}=\hbar /2|$Im$\Sigma(\bm k)_{nn}|$.
	In relatively high magnetic field, the formal Kubo formula is employed to calculate the conductivity, which takes the form
	\be
	\sigma_{\mu\mu}
	=\frac{e^2 \hbar}{\pi V}\text{Tr}\{v^{\mu}(\text{Im}\bar{G})v^{\mu}\text{Im}\bar{G}\}.
	\ee
	Here, $V$ is the volume of the unit cell and  $v^{\mu}$ is the velocity operator along $\mu$-direction. The magnetic field breaks the translational symmetry in real space, which is restored by Peierls substitution with a magnetic supercell in our calculations. When the magnetic field is weak, this Peierls substitution approach requires large supercell that makes the computation intractable. Therefore, in the low-field regime, the conductivity ($\bm B \varparallel\bm E$) is obtained by the semi-classical Boltzmann approach, given by the formula~\cite{Kim2014}
	\be
	\sigma_{\mu\mu}=\sum_{n} e^2 \int \frac{d\bm{k}}{(2\pi)^3} \mathcal D_{n\bk}(v^{\mu}_{n\bk}+\frac{eB^\mu}{\hbar}\bm \Omega_{n\bk}\cdot \bm v_{n\bk})^2 \tau_{nk}
	\delta(\epsilon_{n\bk}-E_F)
	\label{eq:010}
	\ee
	with $\mathcal D_{n\bk} = (1+(e/\hbar)(\bm B\cdot \bm \Omega_{n\bk}))^{-1}$.  $\bm \Omega_{n\bk}$, $\bm{v}_{n\bm k}$, $\epsilon_{n\bk}$, $\tau_{nk}$ are, respectively, the Berry curvature, group velocity, band energy and relaxation time of the $n$th band at wave vector $\bk$. The Landau quantization is neglected in the semi-classical Boltzmann approach.
	
	\emph{Quantum correction to conductivity.---}
	The maximally crossed diagrams are known to be responsible  for the quantum correction\cite{Neal1966}. According to the momentum conservation, each of the diagrams can be represented by their total momentum $\bm q $. Similar to the Drude conductivity, the quantum correction to conductivity under weak magnetic field can not be directly addressed by Peierls substitution. A semi-classical approximation via changing the phase of Green's function is used to handle this situation\cite{wolfle1984electron}, which is believed to be valid in the classic regime ($l_B/l>>1$). $l_B=\sqrt{\hbar/4eB}$ is the magnetic length. The momentum $\bm q^2$ is quantized to  $q_\mu^2+\frac{4eB}{\hbar}(n+1/2)$ with a range from $1/(l_{\phi}^{\mu})^2$ to $1/(l^{\mu})^2$, where $\mu$ labels the direction of the magnetic field. The mean-free path $l^\mu$ along $\mu$-direction relates to the relaxation time by $l^\mu=\sqrt{D_{\mu\mu}\tau}$. $D_{\mu\mu}$ is the diffusion coefficient related to the conductivity via the Einstein relation $\sigma_{\mu\mu}=e^2N(E_F)D_{\mu\mu}$, in which $N(E_F)$ represents the density of states. The coherence length $l_\phi$ along $\mu$-direction charactering the inelastic scattering process is determined by the inelastic relaxation time $\tau_i$ and $D_{\mu\mu}$ with $l_{\phi}^\mu=\sqrt{D_{\mu\mu}\tau_i}$.
	
	\bibliography{refs}

\end{document}